\documentclass[12pt]{iopart}

\makeatletter
\@namedef{ver@amsmath.sty}{}
\makeatother
\usepackage{amstext}
\usepackage{chemformula}

\usepackage[english]{babel}
\usepackage{dcolumn}
\usepackage{bm}
\usepackage{color}
\usepackage{subfigure}
\usepackage{amsfonts}
\usepackage{babel}

\def\bra#1{\langle #1 |}
\def\ket#1{| #1 \rangle}
\def\sinc{\mathop{\text{sinc}}\nolimits}

\def\ee{\mathrm{e}}
\def\dd{\mathrm{d}}
\def\ii{\mathrm{i}}

\def\br{\bm{r}}
\def\bq{\bm{q}}

\newcommand{\dispdot}[2][7mu]{\dot{#2\mkern#1}\mkern-#1}% \dispdot[<disp>]{<stuff>}
\makeatother

\begin{document}

	\title[Spontaneous emission in dispersive media without point-dipole approximation]{Spontaneous emission in dispersive media without point-dipole approximation}
	
	\author{Giovanni Scala$^{1,2}$, Francesco V. Pepe$^{1,2,*}$, Paolo Facchi$^{1,2}$, Saverio Pascazio$^{1,2}$, Karolina S\l owik$^{3}$, }

	\address{$^{1}$ Dipartimento di Fisica and MECENAS, Universit\`{a} di Bari, I-70126 Bari, Italy}
	\address{$^{2}$ INFN, Sezione di Bari, I-70125 Bari, Italy}	
	\address{$^{3}$ Institute of Physics, Nicolaus Copernicus University in Toru\'{n}, Grudziadzka 5, 87-100 Torun, Poland}
	\address{*Corresponding author}
	\ead{francesco.pepe@ba.infn.it}

	\begin{abstract}
We study a two-level quantum system embedded in a dispersive environment and coupled with the electromagnetic field. We expand the theory of light-matter interactions to include the spatial extension of the system, taken into account through its wavefunctions. This is a development beyond the point-dipole approximation. This ingredient enables us to overcome the divergence problem related to the Green tensor propagator. Hence, we can reformulate the expressions for the spontaneous emission rate and the Lamb shift. In particular, the inclusion of the spatial structure of the atomic system clarifies the role of the asymmetry of atomic states with respect to spatial inversion in these quantities.
	\end{abstract}

	\maketitle 
	\section{\label{sec:level1}Introduction} 
The interaction of an atomic system with a surrounding photonic bath yields a correction to the atomic transition energy, referred to as Lamb shift~\cite{Lamb1947}, and gives rise to the process of spontaneous emission. The latter is described in the Markovian limit as an exponential decay~\cite{Einstein1917,Kossakowski2002}, while a much more sophisticated behavior was predicted and verified in non-Markovian regimes~\cite{Crespi,feshbach2020}. If multiple emitters are present, a shared photonic bath acts as a carrier of interactions among them and is responsible for collective emission, such as Casimir effect~\cite{Philbin2010} or Dicke superradiance~\cite{Dicke1954}. For a comprehensive discussion of these and other effects of quantum vacuum on atomic systems see~\cite{Milonni1994}.

The spatial and spectral structure of the photonic bath can be tailored, e.g.\ with traditional cavities or nanostructured materials. As a consequence, the effects arising in atomic systems coupled to such tailored surroundings are modified accordingly~\cite{Hillery1984,Hillery1997}. When it comes to spontaneous emission, this phenomenon has been termed Purcell effect~\cite{purcell1946,Akselrod2014,Zhang2018,Tokman2018}. Similarly, the Lamb shift and collective effects can be tailored by proper engineering of the photonic bath~\cite{Dzsotjan2010, Sinha2018}. In the great majority of works studying light-matter interactions in this context, atomic systems are assumed to be point-like dipoles, without internal structure. This is usually a well-justified approximation, since the size of the atomic system is much below the emission wavelength. However, recent advancement in the field of nanophotonic brought into reach nano- or even picometric cavities \cite{chikkaraddy2016,benz2016}. In the conditions of extreme light confinement, the internal structure of the atomic system might have a considerable impact on its optical response, which might require extensions of the theory beyond the point-dipole ~\cite{stobbe2012,Rivera2016,Neuman2018} or electric-dipole approximations~\cite{Kosik2020}. On the other hand, spatially extended systems like quantum dots may require such treatment even when embedded in a photonic environment as simple as a homogeneous and isotropic medium. 

Accounting for the internal structure of atomic systems can lead to much more than quantitative corrections of their optical properties; actually, effects like spatial asymmetry may give rise to appealing new applications, such as optically-tunable low-frequency radiation sources based on resonantly driven systems \cite{kibis_asymm_semiclass,kibis_asymm_quantum_dot,Chestnov2017}. Scenarios exploiting systems with broken inversion symmetry were proposed for light squeezing~\cite{koppenhofer2016} and lasing~\cite{marthaler2016}. The asymmetry has already been studied in the context of a coherent driving field~\cite{Avetissian2013,Paspalakis2013} with a long list of recent experiments which involve quantum piezoelectricity~\cite{Bimberg2009}, quantum dots~\cite{Bimberg2009}, dye molecules~\cite{Brode1941},
spin-echo~\cite{Stables1998}, Ramsey interferometer~\cite{Martinez-Linares2003}, crystal centers~\cite{Doherty2013,Zhang2018}, and graphene~\cite{Ilani2004,Degen2017}. 

In this work we study the role of the broken inversion symmetry on the spontaneous emission and Lamb shift of an atomic system. We stress how the divergence problem~\cite{Buhmann2012,Hnizdo2004}, encountered when one evaluates the transition properties of atomic systems in dispersive media, can be solved naturally. This problem was treated with many different approaches in other works~\cite{Paknys2016,Scheel1999,Barnett1996,Ho1993}.
Moreover, we exploit the medium-assisted field expressed through the Green tensor propagator~\cite{Qing-HuaQin2007a,Bladel2007}, which can be applied also for classical electrodynamics~\cite{Griffiths2017}, to account for the properties of the photonic surroundings. These can be modified in presence of a host medium, which in general could be structured in terms of geometric shape and spectral response. Although parts of our theory are general, we pay special attention to translationally invariant media. 

The article is organized as follows. In Sec.~\ref{sec:Hamiltonian}, we discuss all the terms of the Hamiltonian; In particular, we describe the form of the coupling between a system of charges and a medium-assisted field, representing in a single entity both the electromagnetic field and the medium charges. In Sec.~\ref{sec:emission}, we use the developed theory to obtain the decay rate and energy shift for an arbitrary bound state, highlighting the contribution of spatial asymmetry of the eigenstates of the atomic Hamiltonian. In Sec.~\ref{sec:examples}, we apply the results to two test-beds. Finally, in Sec.~\ref{sec:conclusions}, we summarize the obtained results and outline future research.

\section{Hamiltonian}\label{sec:Hamiltonian}
	\begin{figure}
	    \centering
	    \includegraphics[scale=0.8]{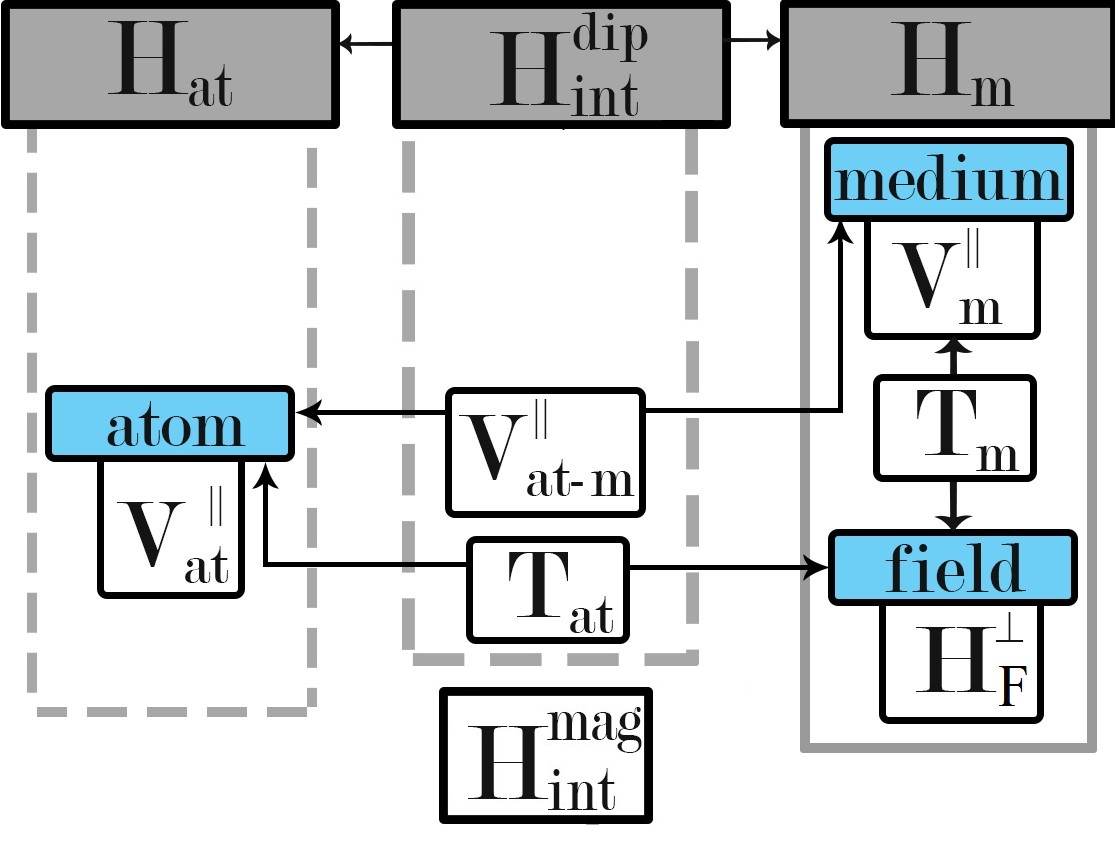}
	    \caption{Graphic view of the total Hamiltonian. In blue, the atomic $V_{\mathrm{at}}^\parallel$, the medium $V_{\mathrm{m}}^\parallel$ [Eq.~\ref{Vparallel}], and the field $H_{\mathrm{f}}^\perp$ [Eq.~\ref{eq:Hamiltonian_free_field}] Hamiltonians. The blue blocks are connected by the interaction blocks. Hence $T_{\textrm{m}}$ and $T_{\textrm{at}}$ connect the medium and the atom, respectively, with the field as in Eq.~\ref{eq:Tminimalcoupling}, and $V_{\textrm{at-m}}^\parallel$ connects the atom and the medium, as in Eq.~\ref{eq:Vqinteract}. These six terms appear in Eq.~\ref{eq:H}. The medium-assisted field $H_{\textrm{m}}$ [Eq.~\ref{eq:HM}] arises from the terms $V_{\mathrm{m}}^\parallel,\,T_{\mathrm{m}}$ and $H_{\mathrm{f}}^\perp$ (solid box on the right), and the atomic Hamiltonian becomes $H_{\mathrm{at}}$ via the PZW transformation [Eq.~\ref{eq:PZW}] (dashed box on the left). $H_{\mathrm{at}}$ interacts with $H_{\mathrm{m}}$ thru $H_{\textrm{int}}$ as in Eq.~\ref{eq:H_summarized}. By neglecting the magnetic properties one obtains $H_{\textrm{int}}^{\textrm{dip}}$, which completes the model investigated. }
	    \label{fig:model}
	\end{figure}
	We start from a first-principle Hamiltonian where positive and negative charges of the atomic system and the medium are coupled with the electromagnetic field. If the focus is on the atomic dynamics, the system can be conveniently modeled by coupling the atom to a \textit{medium-assisted} electromagnetic field, which is dressed by the interaction with the hosting medium (see Fig.~\ref{fig:model}).

Let us consider the Coulomb-gauge Hamiltonian~\cite{Vogel2006,cohen1_1997}, separating the longitudinal and transverse contributions
	\begin{eqnarray}
	H= V_{\text{at}}^{\mathrm{\parallel}}+V_{\mathrm{m}}^{\mathrm{\parallel}}+V_{\text{at-m}}^{\parallel}
	 +T_\mathrm{at}+T_\mathrm{m}+H_\mathrm{F}^\perp.\label{eq:H}
	\end{eqnarray}
	Atomic charges will be labelled by roman indices $j,k$ and the charges of the medium by greek indices $\mu,\nu$. 
The terms
\begin{equation}\label{Vparallel}
    V^\parallel_{\text{at}} = \frac{1}{8\pi\epsilon_{0}}\sum_{j\neq k}\frac{Q_{j}Q_{k}}{|\boldsymbol{r}_{j}-\boldsymbol{r}_{k}|}, \quad V^\parallel_{\text{m}} = \frac{1}{8\pi\epsilon_{0}}\sum_{\mu\neq \nu}\frac{Q_{\mu}Q_{\nu}}{|\boldsymbol{r}_{\mu}-\boldsymbol{r}_{\nu}|}, 
\end{equation} 
represent the internal Coulomb interactions among the charges $Q_k$
of the atomic system (placed at positions $\bm{r}_k$) and among the charges $Q_\mu$ of the medium (placed at positions $\bm{r}_{\mu}$), respectively. The atom-medium Coulomb interactions read
\begin{eqnarray}
	V_{\text{at-m}}^{\parallel} & = \frac{1}{4\pi\epsilon_{0}}\sum_{j,\mu}\frac{Q_{j}Q_{\mu}}{|\boldsymbol{r}_{j}-\boldsymbol{r}_{\mu}|}. \label{eq:Vqinteract}
\end{eqnarray}
The kinetic terms 
\begin{equation}
    T_\mathrm{at} = \sum_{j}\frac{\big(\bm{p}_{j}-Q_{j}\bm{A}(\boldsymbol{r}_{j})\big)^{2}}{2m_{j}},
    \qquad
    T_\mathrm{m} = \sum_{\mu}\frac{\big(\bm{p}_{\mu}-Q_{\mu}\bm{A}(\boldsymbol{r}_{\mu})\big)^{2}}{2m_{\mu}},
    \label{eq:Tminimalcoupling}
\end{equation}
contain the minimal coupling between the charges (with canonical momenta $\bm{p}_j=-\ii\hbar\bm{\nabla}_{\br_j}$ and $\bm{p}_{\mu}=-\ii\hbar\bm{\nabla}_{\bm{r}_{\mu}}$, and masses $m_j$ and $m_\mu$, respectively) and the transverse part of the field, represented by the Coulomb gauge vector potential $\bm{A}$(purely transverse, $\bm{\nabla}\cdot\bm{A}=0$). 
Finally,
\begin{equation}\label{eq:Hamiltonian_free_field}
H_{\text{F}}^{\perp}=\frac{1}{2}\int\mathrm{d}^{3}\boldsymbol{r} \left(\epsilon_0\dispdot{\boldsymbol{A}}^2(\boldsymbol{r})+\frac{1}{\mu_0}\left[\nabla\boldsymbol{A}(\boldsymbol{r}) \right]^2\right)    
\end{equation}
is the Hamiltonian of the free field in vacuum. 
If one considers a neutral atom, the charge density 
\begin{equation}
	\rho_{\mathrm{at}}(\boldsymbol{r})=\sum_{j}Q_{j}\delta(\boldsymbol{r}-\boldsymbol{r}_{j}),\quad\text{with} \quad\sum_{j}Q_{j}=0,
	\end{equation}
can be expressed as the divergence of a polarization density 	
$	\rho_{\mathrm{at}}(\boldsymbol{r})=-\bm{\nabla}\cdot\bm{P}_{\mathrm{at}}(\boldsymbol{r}).
$	
Here,
	\begin{equation}
	\bm{P}_{\mathrm{at}}(\boldsymbol{r})=\sum_{j}q_{j}\int_{0}^{1}\text{d}s\,(\boldsymbol{r}_{j}-\bm{R})\delta(\boldsymbol{r}-\bm{R}-s(\boldsymbol{r}_{j}-\bm{R} )),
	\end{equation}
	where $\bm{R}$ is the center-of-mass coordinate~\cite{cohen1_1997}. The atomic
	polarization density allows us to express the Coulomb interaction terms as follows
	\begin{eqnarray}
	V_{\text{at}}^{\mathrm{\parallel}}= & \frac{1}{2\epsilon_{0}}\int\text{d}^{3}\boldsymbol{r}\left(\bm{P}_{\mathrm{at}}^{\parallel}(\boldsymbol{r})\right)^{2},\label{P_Coulomb}\\
	V_{\text{at-m}}^{\parallel}= & \frac{1}{\epsilon_{0}}\int\text{d}^{3}\boldsymbol{r}\,\bm{P}_{\mathrm{at}}^{\parallel}(\boldsymbol{r})\cdot\bm{\Pi}^{\parallel}(\boldsymbol{r}). \label{Pi_Coulomb}
	\end{eqnarray}
	Here, $\bm{P}_{\mathrm{at}}^{\parallel}$ is the longitudinal part of the polarization, i.e. the only component that determines the atomic charge density, and $\bm{\Pi}^{\parallel}$ is the longitudinal displacement field of the medium, that satisfies
		\begin{equation}
		\nabla\cdot\bm{\Pi}^{\parallel}(\boldsymbol{r})=-\sum_{\mu}Q_{\mu}\delta(\boldsymbol{r}-\boldsymbol{r}_{\mu}).\label{eq:longitudinal_medium_displacement}
		\end{equation}
	The latter is proportional to the Coulomb field $\bm{E}^{\parallel}=-\bm{\Pi}^{\parallel}/\epsilon_0$ generated by the medium charges.

	\subsection{Minimal coupling}
	
	We now analyze the coupling between the atom and the
	electromagnetic field, which is a consequence of the minimal coupling in the kinetic energy terms in Eq.~(\ref{eq:Tminimalcoupling}). For an atom modeled as a point-like dipole, it is possible to shift from the ``$\bm{p}\cdot\bm{A}$'' to the ``$\br\cdot\bm{E}$'' coupling representation, through the unitary transformation $\exp(-\ii Q \br\cdot\bm{A}/\hbar)$, where the vector potential is computed at the dipole center of mass~\cite{cohen1_1997}. The advantage of this transformation lies in the fact that, in the transformed picture, the canonical momentum of a particle coincides with its kinetic momentum and it is decoupled from the field variables (a thorough discussion of the implications of such a feature is given in Ref.~\cite{cohen1_1997}).
	
In the case of a finite-size dipole, the aforementioned unitary transformation generalizes to the Power-Zienau-Wolley (PZW) operator~\cite{cohen1_1997,Vogel2006}:
	\begin{equation}\label{eq:PZW}
\fl U_{\mathrm{PZW}} =  \exp\left(-\frac{\ii}{\hbar}\int\dd^{3}\br\,\bm{P}_{\mathrm{at}}(\br)\cdot\bm{A}(\br)\right)
	 =  \exp\left(-\frac{\ii}{\hbar}\int\dd^{3}\br\,\bm{P}_{\mathrm{at}}^{\perp}(\br)\cdot\bm{A}^{\perp}(\br)\right).
	\end{equation}
The transformation property 
	$
	U_{\mathrm{PZW}}\bm{\Pi}^{\perp}(\br) U_{\mathrm{PZW}}^{\dagger}=\bm{\Pi}^{\perp}(\br)+\bm{P}_{\mathrm{at}}^{\perp}(\br)
$
	yields two transverse-field terms from Eq.~(\ref{eq:Hamiltonian_free_field})
	\begin{equation}
	V_{\mathrm{at}}^{\perp}  =\frac{1}{2\epsilon_{0}}\int\dd^{3}\br\left(\bm{P}_{\mathrm{at}}^{\perp}(\br)\right)^{2},\qquad
	V_{\text{at-m}}^{\perp}=  \frac{1}{\epsilon_{0}}\int\dd^{3}\br\,\bm{P}_{\mathrm{at}}^{\perp}(\br)\cdot\bm{\Pi}^{\perp}(\br).
	\end{equation}
	These contributions are complementary to the ones in Eqs.~(\ref{P_Coulomb}-\ref{Pi_Coulomb}). The latter, as well as the transverse part of the atomic polarization density, are instead
	left unchanged by the transformation. Although originally $\bm{\Pi}^{\perp}=-\epsilon_{0}\bm{E}^{\perp}$, the proportionality is lost after the transformation 
		\begin{equation}
		\bm{\Pi}(\br)=-\epsilon_{0} U_{\mathrm{PZW}}\bm{E}(\br) U_{\mathrm{PZW}}^\dagger-U_{\mathrm{PZW}}\bm{P}_{\text{at}}(\br) U_{\mathrm{PZW}}^\dagger,
		\end{equation}
	which can be shown using Eq.~\ref{eq:longitudinal_medium_displacement}.
	For a finite-size dipole, the equality between the kinetic and canonical
	momenta is not exactly realized in the transformed frame as in the case of a point-like dipole transformation. The reason is that the transformed kinetic momentum 
	\begin{equation}
	\fl
	U_{\mathrm{PZW}}(\bm{p}_{j}+Q_{j}\bm{A}(\br_{j})) U_{\mathrm{PZW}}^{\dagger} =\bm{p}_{j} 
	+Q_{j}\int_{0}^{1}\dd s\,s  (\br_{j}-\bm{R}) \bm{B}(\bm{R}+s(\br_{j}-\bm{R}))
	\end{equation}
	acquires an additional term, which generates a direct coupling between the charges and the magnetic field $\bm{B}$. Nevertheless, the difference between the two momenta in the transformed representation is suppressed with respect to the analogous difference in the Coulomb gauge as the ratio between the atomic size and the interacting light wavelength. Therefore, if one neglects the interaction with the magnetic field, it can be consistently assumed that $\bm{p}_{j}$ coincides with the $j$--th particle kinetic momentum in the transformed representation.
	\subsection{Medium-assisted electromagnetic field}
	\label{sec:medium}
	The medium-assisted electromagnetic field is an effective model that conveniently describes, under certain approximations, the combination of the medium and the field degrees of freedom, as pictured in Fig.~\ref{fig:model}. The contributions to the medium-assisted Hamiltonian arise from the terms $V_{\text{m}}^{\mathrm{\parallel}}$, $T_\mathrm{m}$ and $H_\mathrm{F}^\perp$ in the Hamiltonian~(\ref{eq:H}), as derived in detail in Refs.~\cite{Barnett1996,Huttner1992,Dung2000}. The resulting effective field Hamiltonian,
	\begin{equation}
	H_{\mathrm{m}}=\int_{0}^{\infty}\mathrm{d}\omega\int\mathrm{d}^{3}\boldsymbol{r}\,\hbar\omega\bm{f}^{\dagger}(\bm{r},\omega)\cdot\bm{f}(\bm{r},\omega),\label{eq:HM}
	\end{equation}
	can be expanded in three-component mode operators $\bm{f}(\bm{r},\omega)$ and $\bm{f}^{\dagger}(\bm{r},\omega)$,
	satisfying canonical commutation relations
	\begin{eqnarray}
	\left[f_{k}\left(\bm{r},\omega\right),f_{k'}^{\dagger}\left(\bm{r}',\omega'\right)\right] & =\delta_{kk'}\delta\left(\bm{r}-\bm{r}'\right)\delta\left(\omega-\omega'\right),\label{eq:CCRff1}\\
	\left[f_{k}\left(\bm{r},\omega\right),f_{k'}\left(\bm{r}',\omega'\right)\right] & =\left[f_{k}^{\dagger}\left(\boldsymbol{r},\omega\right),f_{k'}^{\dagger}\left(\bm{r}',\omega'\right)\right]=0,\nonumber
	\end{eqnarray}
 with $k= 1,2,3$.
	
The displacement field \textbf{$\boldsymbol{\Pi}$} and
	the vector potential $\bm{A}$ are related to the field variable
	$\bm{f}$ by
	\begin{equation}
	\fl
	\Pi_{j}(\boldsymbol{r})= \int_{0}^{\infty}\mathrm{d}\omega\int\mathrm{d}^{3}\boldsymbol{r}' 
	\Bigg[-\mathrm{i}\frac{\omega^{2}}{c^{2}}  \sqrt{\frac{\hbar\epsilon_{0}}{\pi}\epsilon_{I}(\boldsymbol{r}',\omega)}G_{jk}(\boldsymbol{r},\boldsymbol{r}',\omega)f_{k}(\boldsymbol{r}',\omega)+\mathrm{H.c.}\Bigg],\label{eq:Pij}\end{equation}
	\begin{equation}
	\fl
	A_{j}(\boldsymbol{r})= \int_{0}^{\infty}\mathrm{d}\omega\int\mathrm{d}^{3}\boldsymbol{r}' 
	\Bigg[\frac{\omega}{c^{2}}  \sqrt{\frac{\hbar}{\pi\epsilon_{0}}\epsilon_{I}(\boldsymbol{r}',\omega)}G_{jk}^{\perp}(\boldsymbol{r},\boldsymbol{r}',\omega)f_{k}(\boldsymbol{r}',\omega)+\mathrm{H.c.}\Bigg], \label{eq:Amedium}
	\end{equation}
	where $\epsilon_{I}$ is the imaginary part of the dielectric permittivity
	\begin{equation}
	\epsilon(\br,\omega) = \epsilon_R(\br,\omega) + \ii \epsilon_I(\br,\omega).
	\end{equation}
	We have assumed that the medium is isotropic, hence the permittivity is a scalar. The Green tensor $G$ appearing in Eq.~(\ref{eq:Pij}) is the solution of the equation~\cite{Vogel2006}
	\begin{equation}
	\left[\partial_{j}\partial_{\ell}-\delta_{j\ell}\left(\nabla^{2}+\frac{\omega^{2}}{c^{2}}\epsilon(\boldsymbol{r},\omega)\right)\right]G_{\ell k}(\boldsymbol{r},\boldsymbol{r}',\omega)
    =\delta_{jk}\delta(\boldsymbol{r}-\boldsymbol{r}'), \label{eq:eqG}
	\end{equation}
	and the term $G^{\perp}$ in Eq.~(\ref{eq:Amedium}) represents its transverse part, satisfying $\partial G_{\ell k}^{\perp}(\boldsymbol{r},\boldsymbol{r}',\omega)/\partial r_{\ell} =\partial G_{k\ell}^{\perp}(\boldsymbol{r}',\boldsymbol{r},\omega) / \partial r_{\ell}'=0$.
	In the Coulomb gauge, the properties of the Green tensor and the analytic structure of $\epsilon(\br,\omega)$ in the complex frequency plane guarantee that the vector potential and the transverse part of the displacement field satisfy the canonical commutation relations
 	\begin{equation}
	\left[A_j(\br),\Pi_k(\br')\right]  = \ii \hbar \delta_{jk}^{\perp} (\br-\br')   = \ii \hbar \int\frac{\dd^3\bq}{(2\pi)^3} \left(\delta_{j\ell}-\frac{q_{j}q_{\ell}}{|\boldsymbol{q}|^{2}}\right) \ee^{\ii\bq\cdot(\br-\br')}.
	\end{equation}
For a translationally invariant medium,  $\epsilon(\boldsymbol{r},\omega)=\epsilon(\omega)$, thus
	the Green tensor depends only on the coordinate difference, $G(\bm{r},\bm{r}',\omega)=G(\bm{r}-\bm{r}',\omega)$, and its Fourier transform 
	\begin{equation}
	\tilde{G}_{jk}(\bm{q},\omega)=\int\mathrm{d}^{3}\boldsymbol{r}\,G_{jk}(\boldsymbol{r},\omega)\mathrm{e}^{-\mathrm{i}\bm{q}\cdot\boldsymbol{r}},
	\end{equation}
reads
	\begin{eqnarray}
	\tilde{G}_{jk}^{\perp}(\boldsymbol{q},\omega)= & \left(\delta_{j\ell}-\frac{q_{j}q_{\ell}}{|\boldsymbol{q}|^{2}}\right)\tilde{G}_{\ell k}(\boldsymbol{q},\omega)=\frac{\delta_{jk}-q_{j}q_{k}/|\boldsymbol{q}|^{2}}{|\boldsymbol{q}|^{2}-\omega^{2}\epsilon(\omega)/c^{2}},\nonumber \\
	\tilde{G}_{jk}^{\parallel}(\boldsymbol{q},\omega)= & \frac{q_{j}q_{\ell}}{|\boldsymbol{q}|^{2}}\tilde{G}_{\ell k}(\boldsymbol{q},\omega)=-\frac{q_{j}q_{k}}{|\boldsymbol{q}|^{2}}\frac{c^{2}}{\omega^{2}\epsilon(\omega)}.\label{eq:G}
	\end{eqnarray}
	Hence, the displacement field reduces to
	\begin{equation}
	\fl
	\Pi_{j}(\boldsymbol{r})=  \int_{0}^{\infty}\mathrm{d}\omega\int\frac{\mathrm{d}^{3}\boldsymbol{q}}{(2\pi)^{3}} 
	  \left[-\mathrm{i}\frac{\omega^{2}}{c^{2}}\sqrt{\frac{\hbar\epsilon_{0}}{\pi}\epsilon_{I}(\omega)}\tilde{G}_{jk}(\boldsymbol{q},\omega)\tilde{f}_{k}(\boldsymbol{q},\omega)\mathrm{e}^{\mathrm{i}\boldsymbol{q}\cdot\boldsymbol{r}}+\mathrm{H.c.}\right],
	\end{equation}
	where the operators
	\begin{equation}
	\tilde{\bm{f}}(\boldsymbol{q},\omega)=\int\mathrm{d}^{3}\boldsymbol{r}\,\bm{f}(\boldsymbol{r},\omega)\mathrm{e}^{-\mathrm{i}\bm{q}\cdot\boldsymbol{r}},
	\end{equation}
	satisfy 
	\begin{equation}
	[\tilde{f}_{j}(\boldsymbol{q},\omega),\tilde{f}_{k}^{\dagger}(\boldsymbol{q}',\omega')]=(2\pi)^{3}\delta_{jk}\delta(\omega-\omega')\delta(\boldsymbol{q}-\boldsymbol{q}').
	\end{equation}
	For a point-like atomic system, singularities may arise in the interaction
	Hamiltonian due to the fact that the quantities
	$G^{\parallel}(\bm{r},\omega)$ and $G^{\perp}(\bm{r},\omega)$ diverge as $\bm{r}\to\bm{0}$.
	In fact, while 
	\begin{equation}
	\fl
	\mathrm{Im} G^{\perp}_{jk}(\bm{0},\omega) = \int \frac{\dd^3\bq}{(2\pi)^3} \mathrm{Im} \tilde{G}^{\perp}_{jk}(\bq,\omega)  
	 = \frac{\omega^2\epsilon_I(\omega)}{c^2} \int \frac{\dd^3\bq}{(2\pi)^3} \frac{\delta_{jk}-q_{j}q_{k}/|\boldsymbol{q}|^{2}}{\left||\boldsymbol{q}|^{2}-\frac{\omega^2\epsilon(\omega)}{c^2}\right|^2}
	\end{equation}
	is finite and yields a well-defined transverse decay rate~\cite{Barnett1996}, $\mathrm{Im} G^{\parallel}(\bm{r},\omega)$ diverges as $\bm{r}\to\bm{0}$, due to the non integrability of $\mathrm{Im} \tilde{G}_{jk}^{\parallel}(\bq,\omega)\propto q_jq_k/|\bq|^2$, and a consistent treatment of the longitudinal decay rate requires momentum regularization. 
	
	Techniques based on considering the source enclosed in an artificial cavity~\cite{ Juzeliunas1997,Scheel1999,Barnett1999} have been developed to cope with such singularities. In the following, we will tackle the divergences of the longitudinal part with a less artificial approach, by considering the natural finite spatial extent of the atomic wavefunctions.
This will allow us to unambiguously analyze the role of the asymmetry of the atomic states on the emission process.

	\subsection{Total Hamiltonian}
	From the previous parts of this section it follows that
	\begin{equation}
	H=H_{\mathrm{at}}+H_{\mathrm{int}}^{\mathrm{el}}+H_{\mathrm{int}}^{\mathrm{mag}}+H_{\mathrm{m}}.\label{eq:H_summarized}
	\end{equation}
	Here,
	\begin{equation}
	H_{\mathrm{at}}=H^{0}_{\mathrm{at}}+V_{\mathrm{at}}=\sum_{j}\frac{\bm{p}_{j}^{2}}{2m_{j}}+\frac{1}{2\epsilon_{0}}\int\dd^{3}\br\left(\bm{P}_{\mathrm{at}}(\br)\right)^{2}
	\end{equation}
	is the atomic Hamiltonian, 
	\begin{equation*}
		\fl
	H_{\mathrm{int}}^{\mathrm{el}}=  \frac{1}{\epsilon_{0}}\int\dd^{3}\br\,\bm{P}_{\mathrm{at}}(\br)\cdot\bm{\Pi}(\br)
	=  \frac{1}{\epsilon_{0}}\sum_{j}Q_{j}(\br_{j}-\bm{R})\cdot\int\dd s\,\bm{\Pi}(\bm{R}+s(\br_{j}-\bm{R}))
	\end{equation*}
	represents the interaction of the atomic system with the electric field, and
	\begin{eqnarray}
	H_{\mathrm{int}}^{\mathrm{mag}}=  \sum_{j}&\Bigg\{\frac{Q_{j}}{m_{j}}\bm{p}_{j}\cdot\int_{0}^{1}\dd s\,
	 s(\br_{j}-\bm{R})\bm{B}(\bm{R}+s(\br_{j}-\bm{R}))\nonumber\\
	&+\frac{Q_{j}^{2}}{2m_{j}} \left[\int_{0}^{1}\dd s\,s(\br_{j}-\bm{R})\bm{B}(\bm{R}+s(\br_{j}-\bm{R}))\right]^{2}\Bigg\} 
	\end{eqnarray}
	stands for the coupling with the magnetic field. The term $H_{\mathrm{m}}$ generally represents the Hamiltonian of the medium, that can be modeled in different ways, e.g.\ through the medium-assisted field Hamiltonian~(\ref{eq:HM}), as shown in Sect.~\ref{sec:medium}.

	In the following part of this work we will neglect the magnetic contribution to the interaction. We will model the atom as a dipole of charge $Q$, with a heavy positive
	charge at the fixed position $\bm{R}=0$ and a moving negative charge
	of coordinate $-\br$ and mass $m$. As a result, one finds the final form of the interaction Hamiltonian
	\begin{equation}
	H_{\mathrm{int}}^{\mathrm{dip}}=\frac{Q}{\epsilon_{0}}\br\cdot\int_{0}^{1}\dd s\,\bm{\Pi}(-s\br),
	\end{equation}
	representing the correct generalization of the ``$\br\cdot\bm{E}$'' Hamiltonian to an extended (non-pointlike) dipole. This is the main finding of this work, that arises as a connection between first-principle QED, represented through the canonical commutation relations, and the medium-assisted field ruled by Eq.~(\ref{eq:CCRff1}).

	\section{Emission properties of a bound system of charges} \label{sec:emission}
	According to the results of the previous section, each eigenstate of the internal atomic Hamiltonian is dressed by the surrounding medium. 
	We now characterize the single-photon
	emission process and the Lamb shift of an atomic level in a medium-assisted photonic environment in a translationally invariant medium. 
	
	Consider an atom in an arbitrary environment, i.e. a dispersive medium of any geometry and material. Let $\ket{a}$ and $\ket{b}$ be two orthogonal eigenstates of the free atomic Hamiltonian $H_{\mathrm{at}}$, characterized by
	\begin{equation}
	H_{\mathrm{at}}\ket{a} = E_a \ket{a}, \quad H_{\mathrm{at}}\ket{b} = E_b \ket{b} .
	\end{equation} 
	The atom-photon interaction is described by the matrix element 
	\begin{equation}\label{matrixM}
	    \mathcal{M}_{j}^{ab}(\br,\omega)=\bra{a}H_{\mathrm{int}}^{\mathrm{dip}}f_{j}^{\dagger}(\br,\omega)\ket{b},
	\end{equation}
	which, for a translationally-invariant medium,  can be expressed in the Fourier space through
	\begin{equation}
	\fl
	\tilde{\mathcal{M}}_{j}^{ab}  (\bq,\omega) =  \bra{a}H_{\mathrm{int}}^{\mathrm{dip}}\tilde{f}_{j}^{\dagger}(\bm{q},\omega)\ket{b}
	=-\ii C(\omega) \frac{\omega^2}{c^2}
	\tilde{G}_{jk}(\bq,\omega)  \bra{a}r_{k}\int_{0}^{1}ds \; \ee^{-\ii s\bq\cdot\br}\ket{b},\label{eq:M} 
	\end{equation}
	where 
	\begin{equation*}
	   C(\omega)= Q\sqrt{\frac{\hbar\epsilon_{I}(\omega)}{8\pi^{4}\epsilon_{0}}}.
	\end{equation*}
    If we insert the expression of $\tilde{G}_{jk}$ in Eq.~(\ref{eq:G}) and exploit the orthogonality between longitudinal and transverse projectors, we obtain
	\begin{eqnarray}
	\mathcal{T}_{ab}(\bq,\omega) &=&  \sum_{j=1}^{3}\left|\tilde{\mathcal{M}}_{j}^{ab}(\bq,\omega)\right|^{2}\\&=&  \frac{C(\omega)^2}{|\epsilon(\omega)|^2} \Biggl[ \mathcal{D}(|\bq|,\omega)\mathcal{G}_{ab}(\bq)  + (1-\mathcal{D}(|\bq|,\omega)) \frac{|\mathcal{F}_{ab}(\bq)-\delta_{ab}|^2}{|\bq|^2} \Biggr],\nonumber  \label{eq:T}
	\end{eqnarray}
	where $\delta_{ab}=\langle a|b \rangle=1$ if $\ket{a}$ and $\ket{b}$ coincide and 0 otherwise, with 
	\begin{eqnarray}
	\mathcal{D}(q,\omega) & = \left| 1- \frac{q^2 c^2}{\omega^2 \epsilon(\omega)} \right|^{-2}, \\
	\mathcal{F}_{ab}(\bm{q})&=\bra{a}\ee^{-\ii\bq\cdot\br}\ket{b}=\int\dd^{3}\br\,\psi_{a}^{*}(\br)\psi_{b}(\br)\ee^{-\ii\bq\cdot\br}, \\
	\mathcal{G}_{ab}(\bq) & = \sum_{j=1}^3 \left| \bra{a} r_j \frac{\ee^{-\ii\bq\cdot\br}-1}{\bq\cdot\br} \ket{b} \right|^2 .
	\end{eqnarray}
	The quantity defined in Eq.~(\ref{eq:T}) determines both the total decay rate of the state $\ket{a}$ and its energy shift. The former can be evaluated according to the Fermi golden rule
	\begin{equation}
	\Gamma_{a}  =\frac{2\pi}{\hbar}\sum_{b}\int_{0}^{\infty}\dd\omega\delta(\hbar\omega-\hbar\omega_{ab}) T_{ab}(\omega) 
	=\frac{2\pi}{\hbar^{2}}\sum_{b\neq a}\theta(\omega_{ab}) T_{ab}(\omega_{ab}),\label{eq:Gamma}
	\end{equation}
	with
	\begin{equation}
	\omega_{ab}  = \frac{E_a-E_b}{\hbar}, \qquad
	T_{ab}(\omega)  = \int\dd^{3}\bq\,\mathcal{T}_{ab}(\bq,\omega) , \label{eq:Tab}
	\end{equation}
and $\theta(x)$ being the Heaviside step function. 
	The absence of a contribution from state $\ket{a}$ in the sum over states in the second equality of Eq.~(\ref{eq:Gamma}), albeit reasonable, is not a trivial result. Therefore, in the evaluation of the decay rate $\delta_{ab}=0$ and the apparent divergence in the term proportional to $\mathcal{F}_{ab}$ is regularized by the wavefunctions spatial extension. 
	
	In vacuum ($\epsilon(\omega)=1$), the decay rate in Eq.~\ref{eq:Gamma} becomes
	\begin{equation}
	    \Gamma_a^{\mathrm{(vac)}}=\frac{Q^2 q^3}{8\pi^2\hbar\epsilon_0}\int_{\mathbb{S}^2}
	    \dd^2 S(\bm{n})  \sum_{b\neq a}\Big[ \mathcal{G}_{ab}(q\bm{n})-\frac{|\mathcal{F}_{ab}(q\bm{n})|^2} {q^2} \Big],
	\end{equation}
where 
$q= \omega_{ab} /c$ and the integration is over the unit sphere $\bm{n}\in\mathbb{S}^2$. 
	Note that in the point-dipole limit the quantity $\mathcal{F}_{ab}$ tends to $\delta_{ab}$. In this way, we recover the familiar Weisskopf-Wigner result~\cite{cohen2_1998}. 
	
	The frequency shifts of the atomic levels should be determined using Eqs.~(\ref{eq:T}-\ref{eq:Tab}), through
	\begin{equation}\label{eq:delta_a}
	\Delta_a = \frac{1}{\hbar^2} \sum_{b} \mathrm{P}\!\!\int_{0}^{\infty}\dd\omega\frac{T_{ab}(\omega)}{\omega-\omega_{ab}} ,
	\end{equation}
	with $\mathrm{P}\!\!\int$ denoting principal value integration. 
	For $a=b$ the function $\mathcal{T}_{ab}$ contains the state-independent, non-integrable term
	\begin{equation}
	\delta_{ab} C(\omega)^2 |\bq|^{-2} (1-\mathcal{D}(|\bq|,\omega)) \sim |\bq|^{-2} \quad \text{as } |\bq|\to\infty, \label{eq:nonint}
	\end{equation}
	which provides a divergent contribution to $T_{aa}(\omega)$. However, this contribution is also independent of the state, representing therefore the effect of a uniform energy shift.
	Physical quantities such as the perturbed transition frequency
	\begin{eqnarray}\label{eq:omega_ab}
	\tilde{\omega}_{ab} =  \omega_{ab} + \Delta_a - \Delta_b &
	= \omega_{ab} + \frac{1}{\hbar^2} \mathrm{P}\!\!\int_{0}^{\infty}\dd\omega \Biggl[  \frac{T_{aa}(\omega)-T_{bb}(\omega)}{\omega}  \nonumber \\ & + 2 \omega_{ab} \frac{T_{ab}(\omega)}{\omega^2-\omega_{ab}^2} + \sum_{c\neq a,b} \left( \frac{T_{ac}(\omega)}{\omega-\omega_{ac}} - \frac{T_{bc}(\omega)}{\omega-\omega_{bc}}  \right) \Biggr]
	\end{eqnarray}
	are thus independent of the divergent term given in~(\ref{eq:nonint}). Indeed, notice that, in the time domain the low-energy behavior of the dielectric permittivity is 
	\begin{equation}
	\epsilon(\omega) = 1 + \int_0^{\infty} \dd t \chi(t) + \ii \omega \int_0^{\infty} \dd t \, t\chi(t) + O(\omega^2),
	\end{equation}
	where $\chi(t)$ is the medium susceptibility with finite moments. This implies that $T_{ab}(\omega)\sim\omega$ close to the origin, and therefore the integration of the term $(T_{aa}-T_{bb})/\omega$ in Eq.~(\ref{eq:omega_ab}) is well defined.

	\subsection{Asymmetric two-level atom}
	
	The parity asymmetry of the atomic Hamiltonian eigenstates, reflected by the presence of nonvanishing expectation values of one or more components of $\br$, affects the state-dependent quantities $\mathcal{F}_{ab}$ and $\mathcal{G}_{ab}$, which appear in the expression of $T_{ab}(\omega)$ and determine the decay rate $\Gamma_a$ and the energy shift $\Delta_a$. In a two-level atomic system, the three components of the Hermitian position operator $\bm{r}$ can be represented by spin operators~\cite{feynman1964,cohen2_1998}
	\begin{equation}\label{eq:rspin}
	\bm{r} = \bm{\rho} \bm{1} + \bm{\delta} \sigma_z + \bm{r}_{ab}\sigma_x , \qquad
	\sigma_x = \ket{a}\bra{b}+\ket{b}\bra{a}, \quad \sigma_z = \ket{a}\bra{a}-\ket{b}\bra{b}
	\end{equation}
    acting on the two-dimensional space spanned by $\ket{a},\ket{b}$,
	with 
	\begin{eqnarray}
	\bm{\rho} & = \frac{\bra{a}\bm{r}\ket{a}+\bra{b}\bm{r}\ket{b}}{2} ,\label{eq:rho} \\ \bm{\delta} & = \frac{\bra{a}\bm{r}\ket{a}-\bra{b}\bm{r}\ket{b}}{2} , \\ \bm{r}_{ab} & = \bra{a}\bm{r}\ket{b} = \bra{b}\bm{r}\ket{a} .\label{eq:rab}
	\end{eqnarray}
	In the two-level case, the off-diagonal matrix element~(\ref{eq:rab}) can be made real and non-negative by absorbing a phase factor in the definition of one of the states. 
	
	The functions that determine the decay rate from $\ket{a}$ to $\ket{b}$ read
	\begin{eqnarray}
	\mathcal{F}_{ab}(\bq) & = -\ii \ee^{-\ii \bq\cdot \bm{\rho}} \bq\cdot \br_{ab}\, \mathrm{sinc}(A(\bq)) , \\
	 \mathcal{G}_{ab}(\bq) & =  \left| \nabla_{\bm{q}}\left[ \bq\cdot \br_{ab} \int_0^1 \dd s \sinc(s A(\bq))  \ee^{-\ii s \bq\cdot \bm{\rho}} \right] \right|^2,\nonumber
	\end{eqnarray}
	with $\sinc(x)=\sin(x)/x$
	and
	$
	A(\bq) = \sqrt{(\bq\cdot \br_{ab})^2 + (\bq \cdot \bm{\delta})^2} .
	$
	
	From these results, one can observe that the physical quantities computed from $\mathcal{G}_{ab}$ and from the square modulus of $\mathcal{F}_{ab}$ generally depend on both $\bm{\rho}$ or $\bm{\delta}$, but are invariant with respect to the inversions $\bm{\rho}\to -\bm{\rho}$ and $\bm{\delta}\to\bm{-\delta}$. 
	
	To identify the lowest-order contributions to the decay rate, let us perform a small-$\bq$ expansion of the functions appearing in the expression~(\ref{eq:T}) of $\mathcal{T}_{ab}$ for $a\neq b$, namely
	\begin{equation}\label{eq:Fabappr}
	\frac{|\mathcal{F}_{ab}(\bq)|^2}{|\bq|^2} \simeq \frac{(\bq\cdot\br_{ab})^2}{|\bq|^2} \!\left( 1 - \frac{(\bq\cdot\br_{ab})^2+(\bq\cdot\bm{\delta})^2}{6} \right)\!  ,  
	\end{equation}
	and
	\begin{eqnarray} 
	\mathcal{G}_{ab}(\bq) \simeq & |\br_{ab}|^2 \!\left( 1 - \frac{(\bq\cdot\br_{ab})^2}{3}+\frac{(\bq\cdot\bm{\delta})^2}{9} + \frac{(\bq\cdot\bm{\rho})^2}{12}  \right) \nonumber \\
	& + \frac{|\bm{\rho}|^2 (\bq\cdot\br_{ab})^2}{4} + (\bq\cdot\br_{ab})\br_{ab} \cdot \left( \frac{(\bq\cdot\bm{\rho})\bm{\rho}}{2} - \frac{(\bq\cdot\bm{\delta})\bm{\delta}}{9} \right) . \label{eq:Gabappr}
	\end{eqnarray}
	The second-order approximation in $\bq$ of the functions in Eqs.~(\ref{eq:Fabappr}-\ref{eq:Gabappr}) yield divergent integrals, that should be regularized by a cutoff $\Lambda_{\bq}$, roughly corresponding to the inverse spatial size of the involved wavefunctions, that can range from 1 to 100 nm according to the considered system. In this way, one can estimate that the corrections entailed by an asymmetry of the states $\ket{a}$ and $\ket{b}$ are of order $(\Lambda_{\bq}|\br_{aa}|)^2$ and $(\Lambda_{\bq}|\br_{bb}|)^2$. Notice that the asymmetry corrections compete with terms of order $(\Lambda_{\bm{q}}|\bm{r}_{ab}|)^2$, representing the first corrections to the point-dipole result, and are not characterized by a definite sign.

	\section{Test beds}\label{sec:examples}
	
	In this section, we apply the theory to two systems: a hydrogen atom in a static electric field and an asymmetric quantum well. We shall focus on the dependence of spontaneous emission on their spatial asymmetry and on the embedding in an absorptive medium. 
	
	\subsection{Hydrogen atom in a static electric field}\label{sec:hydrogen_emission}
	
The first example we consider is a hydrogen atom embedded in a homogeneous medium.  
The asymmetry of this system is related to the presence of a static electric field $\mathcal{E}$, whose polarization defines the quantization axis. The asymmetry can be classically explained by a shift of the electronic cloud with respect to the nucleus. As a result, the eigenstates of the system perturbed by the field correspond to superpositions of wavefunctions
\begin{equation}\label{eq:exp1}
 \ket{\psi(\mathcal{E})}=\sum_{nlms}b_{nlms}\left(\mathcal{E}\right)\ket{\psi_{nlm}}\otimes\ket{\chi_s}
\end{equation}
of a bare hydrogen atom, where the orbital wavefunction $\ket{\psi_{nlm}}$ is characterized by the principal ($n$), angular ($l$) and magnetic ($m$) quantum numbers, and $\ket{\chi_s}$ represent the spin up (down) state for $s=+$ ($s=-$). Equivalently, the same state can be decomposed in the Clebsch-Gordan basis 
\begin{equation}\label{eq:exp2}
 \ket{\psi(\mathcal{E})}=\sum_{nljm_j}c_{nljm_j}\left(\mathcal{E}\right)\ket{\phi_{nljm_j}} ,
\end{equation}
with $j$ the total angular momentum and $m_j$ its projection on the third axis. Clebsch-Gordan states corresponding to $n=1,2$, on which the following analysis will be focused, read 
\begin{eqnarray}
\ket{\phi_{10\frac{1}{2}\frac{1}{2}}} &=& \ket{\psi_{100}} \otimes \ket{\chi_+}, \\
\ket{\phi_{10\frac{1}{2}\frac{-1}{2}}} &=& \ket{\psi_{100}} \otimes \ket{\chi_-}, \nonumber \\
\ket{\phi_{20\frac{1}{2}\frac{1}{2}}} &=& \ket{\psi_{200}} \otimes \ket{\chi_+}, \nonumber \\ 
\ket{\phi_{20\frac{1}{2}\frac{-1}{2}}} &=& \ket{\psi_{200}} \otimes \ket{\chi_-}, \nonumber \\
\ket{\phi_{21\frac{1}{2}\frac{1}{2}}} &=& \sqrt{\frac{2}{3}}\ket{\psi_{211}} \otimes \ket{\chi_-} - \sqrt{\frac{1}{3}} \ket{\psi_{210}} \otimes \ket{\chi_+},\nonumber\\
\ket{\phi_{21\frac{1}{2}\frac{-1}{2}}} &=& -\sqrt{\frac{2}{3}} \ket{\psi_{211}} \otimes \ket{\chi_+} + \sqrt{\frac{1}{3}} \ket{\psi_{210}} \otimes \ket{\chi_-},\nonumber \\
\ket{\phi_{21\frac{3}{2}\frac{3}{2}}} &=& \ket{\psi_{211}} \otimes \ket{\chi_+},\nonumber \\
\ket{\phi_{21\frac{3}{2}\frac{1}{2}}} &=& \sqrt{\frac{1}{3}} \ket{\psi_{211}} \otimes \ket{\chi_-} + \sqrt{\frac{2}{3}} \ket{\psi_{210}} \otimes \ket{\chi_+}, \nonumber \\
\ket{\phi_{21\frac{3}{2}\frac{-1}{2}}} &=& \sqrt{\frac{1}{3}} \ket{\psi_{21-1}} \otimes \ket{\chi_+} + \sqrt{\frac{2}{3}} \ket{\psi_{210}} \otimes \ket{\chi_-}, \nonumber \\
\ket{\phi_{21\frac{3}{2}\frac{-3}{2}}} &=& \ket{\psi_{21-1}} \otimes \ket{\chi_-},\nonumber
\end{eqnarray}
Notice that states $\ket{\phi_{n0jm_j}}$ and $\ket{\phi_{n1jm_j}}$ correspond, in the spectroscopic notation, to $ns_{j,m_j}$ and $np_{j,m_j}$, respectively. In our analysis we will adapt the discussion from Ref.~\cite{CohenQM} to the case of electric fields weak enough to see its gradual influence on the eigenstates. As a consequence, the expansion coefficients depend on the applied field as suggested above in Eqs.~(\ref{eq:exp1})-(\ref{eq:exp2}). This result is achieved if the corrections induced by the field are small with respect to the fine structure, and comparable with the Lamb shift. In the opposite case of fields strong enough to overcome the fine structure, the eigenstates are fixed and only their energies still depend on the field.

We will now identify the eigenstates in the weak-field regime, and discuss the evaluation of the transition rate between a selected pair of these eigenstates. As anticipated, we restrict the analysis to the $n=1,2$ manifolds and neglect the small impact of states with $n>2$. If one neglects fine-structure splitting and Lamb shift, the eigenenergies of the $n=1$ and $n=2$ sectors can be set to $\epsilon_1=-13.6\left(1-\frac{1}{2^2}\right)\,\mathrm{eV} = - 10.2 \,\mathrm{eV} $ and $\epsilon_2=0$.
The Hamiltonian $H_0$, restricted to the sector spanned by the Clebsch-Gordan basis, ordered as above, is diagonal in the absence of the field, while, in the general case, it reads 
\begin{eqnarray}
\fl H_0=&\left( 
 \begin{array}{cccccccccc}
  \epsilon_1 & 0 & 0 & 0 & -b_1\mathcal{V} & 0 & 0 & b_2\mathcal{V} & 0 & 0 \\
  0 & \epsilon_1 & 0 & 0 & 0 &  b_1\mathcal{V} & 0 & 0 & b_2\mathcal{V} & 0 \\
  0 & 0 & \Delta_L & 0 & \sqrt{3}\mathcal{V} & 0 & 0 & \sqrt{6}\mathcal{V} & 0 & 0\\
  0 & 0 & 0 & \Delta_L & 0 & -\sqrt{3}\mathcal{V} & 0 & 0 & \sqrt{6}\mathcal{V} & 0 \\
  -b_1\mathcal{V} & 0 & \sqrt{3}\mathcal{V} & 0 & 0 & 0 & 0 & 0 & 0 & 0 \\
  0 & b_1\mathcal{V} & 0 & -\sqrt{3}\mathcal{V} & 0 & 0 & 0 & 0 & 0 & 0 \\
  0 & 0 & 0 & 0 & 0 & 0 & \Delta_{FS} & 0 & 0 & 0 \\
  b_2\mathcal{V} & 0 & \sqrt{6}\mathcal{V} & 0 & 0 & 0 & 0 & \Delta_{FS} & 0 & 0 \\
  0 & b_2\mathcal{V} & 0 & \sqrt{6}\mathcal{V} & 0 & 0 & 0 & 0 & \Delta_{FS} & 0 \\
  0 & 0 & 0 & 0 & 0 & 0 & 0 & 0 & 0 & \Delta_{FS} \nonumber
 \end{array}
 \right) \label{eq:hydrogen_hamiltonian}
\end{eqnarray}
Here $\Delta_{FS}=44\,\mu\mathrm{eV}$ and $\Delta_{L}=4.4\,\mu\mathrm{eV}$ represent respectively the fine structure splitting and the Lamb shift for hydrogen, and $\mathcal{V}=\mathcal{E}ea_0$, with $e$ the elementary charge and $a_0$ the Bohr radius. The constants $b_1=\frac{128}{243}\sqrt{\frac{2}{3}}$ and $b_2=\frac{256}{243}\sqrt{\frac{1}{3}}$ and other off-diagonal elements can be evaluated through an explicit calculation of the corrections $-\mathcal{E}e\langle \phi_{10jm_j}|z| \phi_{21j^\prime,m_{j^\prime}}\rangle$.

\begin{figure}\centering
 \includegraphics[width=0.65\textwidth]{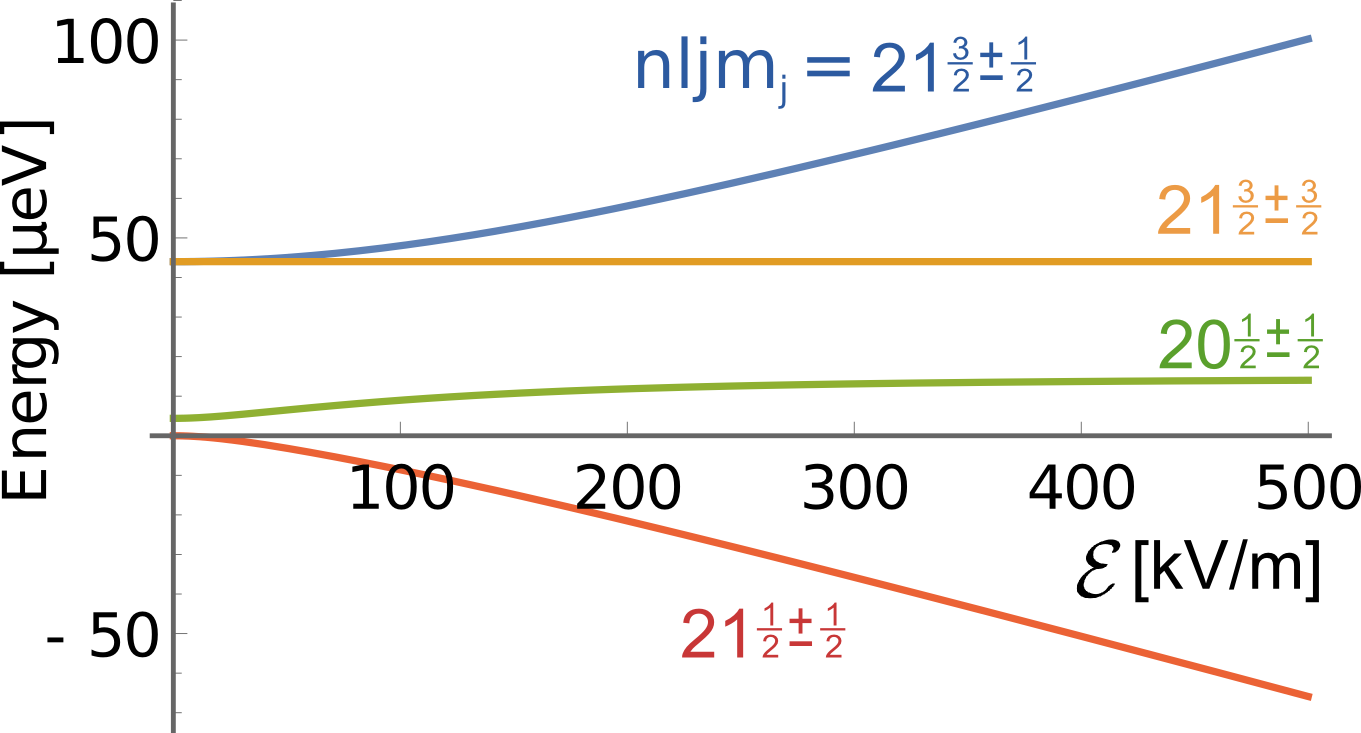}
 \caption{Energies of hydrogen atom eigenstates as functions of a static electric field. Labels, in the spectroscopic notation, are referred to the dominant contribution for $\mathcal{E}\to 0$. As the field strength increases, the label states mixed with orthogonal Clebsch-Gordan states (see Fig.~\ref{fig:hydrogen_expansion}). Each line corresponds to a pair of states with fixed $|m_j|$.\label{fig:hydrogen_energies}}
\end{figure}
Diagonalizing the above Hamiltonian, we find the eigenstates of the system. Our first observation is that the eigenstates originating at the $n=1$ manifold are barely distorted by the field, and their energy is shifted by a correction of the order of peV. In the following analysis we neglect these corrections, both in the eigenstate and in its energy. The dependence of eigenenergies of the $n=2$ manifold on the field is shown in Fig.~\ref{fig:hydrogen_energies}, and again the influence of states from the $n=1$ manifold is negligible. For this reason, from now on we consider the Hamiltonian (\ref{eq:hydrogen_hamiltonian}) with $b_1=b_2=0$. 

\begin{figure}\centering
 \includegraphics[width=0.75\textwidth]{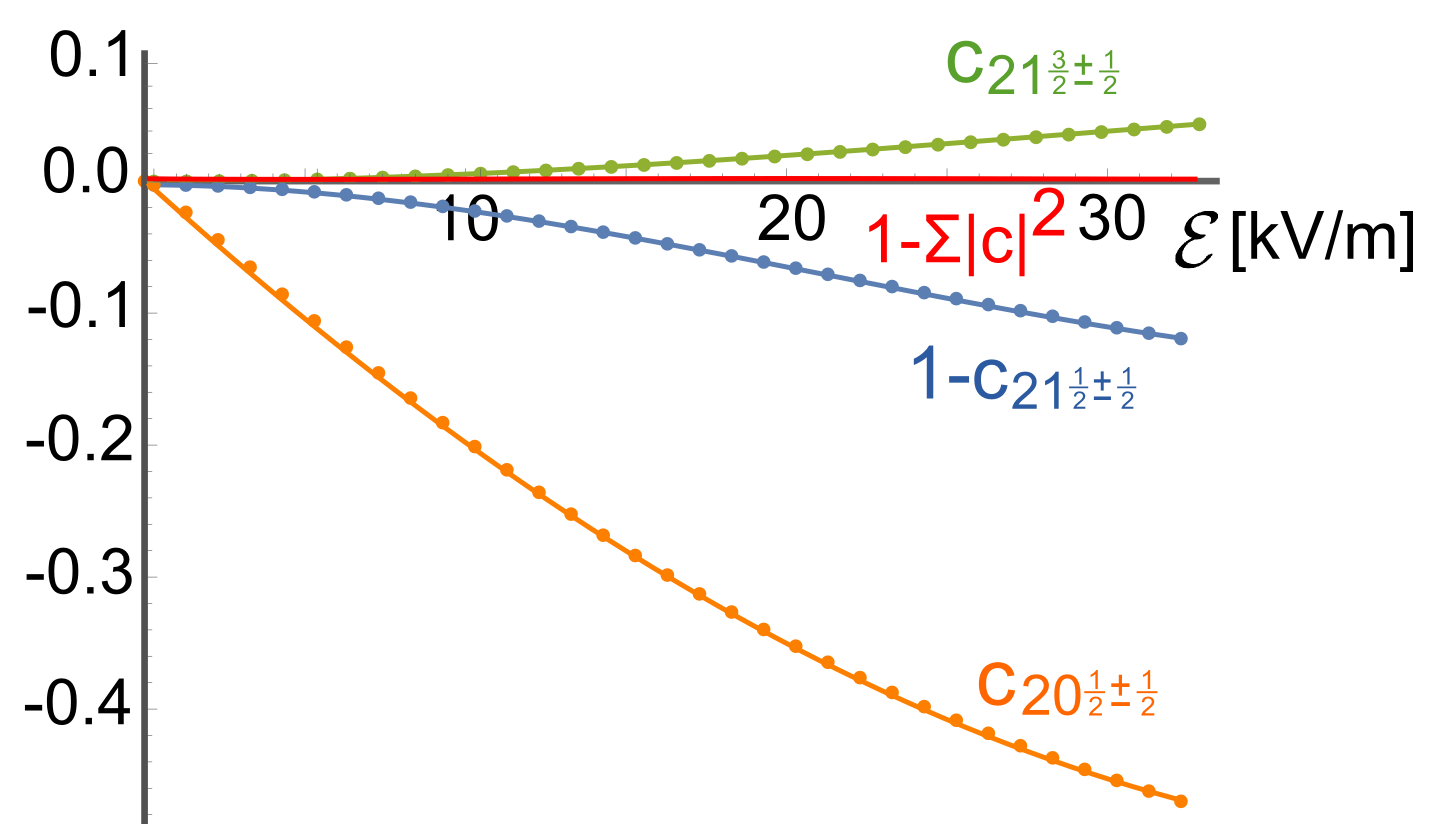}
 \caption{Expansion coefficients of the lowest-excited states (red line in Fig.~\ref{fig:hydrogen_energies}) in terms of Clebsch-Gordan states [see Eq.(\ref{eq:exp2})]. The dots correspond to numerical solutions for selected values of field $\mathcal{E}$. The solid line is a third-order polynomial fit. The red line corresponds to the sum of squares of the three coefficients shown in the figure in blue, orange and green. The sum differs from 1 by less than 0.0005 for the studied range of fields. \label{fig:hydrogen_expansion}}
\end{figure}

From Fig.~\ref{fig:hydrogen_energies} it is clear that the pair of lowest-excited states corresponds to the red line and simplifies to the states $2p_{\frac{1}{2},|m_j|=\frac{1}{2}}$ in the absence of the field. The explicit expansion of these eigenstates in terms of Clebsch-Gordan states and in function of the field is cumbersome. Instead, we find the expansion coefficients numerically and fit them with third-order polynomial functions of $\mathcal{E}$ (Fig.~\ref{fig:hydrogen_expansion}).
For positive $\mathcal{E}$, the expansion coefficients are
\begin{eqnarray}
\fl \ket{\psi_{\mathrm{le},\,m_j}} &=&  c_{21\frac{1}{2}m_j}\ket{\phi_{21\frac{1}{2}m_j}} + c_{20\frac{1}{2}m_j} \ket{\phi_{20\frac{1}{2}m_j}} + c_{21\frac{3}{2}m_j} \ket{\phi_{21\frac{3}{2}m_j}}, \quad (\text{with } m_j=\pm 1/2)\\
\fl c_{21\frac{1}{2}m_j}(\mathcal{E}) & \approx & 1-1.28\times10^{-7}\mathrm{eV}^{-1}\mathcal{E}-2.20\times10^{-10}\mathrm{eV}^{-2}\mathcal{E}^2+3.49\times10^{-15}\mathrm{eV}^{-3}\mathcal{E}^3\nonumber\\
\fl c_{20\frac{1}{2}m_j}(\mathcal{E}) & \approx & -2.20\times10^{-5}\mathrm{eV}^{-1}\mathcal{E}+2.17\times10^{-10}\mathrm{eV}^{-2}\mathcal{E}^2+5.10\times10^{-16}\mathrm{eV}^{-3}\mathcal{E}^3\nonumber\\
\fl c_{21\frac{3}{2}m_j}(\mathcal{E}) & \approx & 1.19\times10^{-9}\mathrm{eV}^{-1}\mathcal{E}+6.58\times10^{-11}\mathrm{eV}^{-2}\mathcal{E}^2-7.63\times10^{-16}\mathrm{eV}^{-3}\mathcal{E}^3\nonumber
\end{eqnarray}
where the subscript ``le'' stands for ``lowest-excited''. With the third-order expansion, the state is normalized to 1 with error smaller than $0.05\%$ for $\mathcal{E}<35$ keV/m. 

There are four possible transitions between a doubly-degenerate excited and a doubly-degenerate ground state. We now select two example transitions among them, namely i) the transition between the excited and ground states with $m_j=-\frac{1}{2}$
\begin{eqnarray}\label{eq:psi_hydrogen}
\ket{\psi_a} &=& \ket{\psi_{\mathrm{le},\,m_j=-\frac{1}{2}}} =
         b_{200\frac{-1}{2}}(\mathcal{E}) \ket{\psi_{200}}\otimes\ket{\chi_-} + b_{210\frac{-1}{2}} (\mathcal{E}) \ket{\psi_{210}}\otimes\ket{\chi_-} \nonumber \\
				& & + b_{211\frac{1}{2}} (\mathcal{E}) \ket{\psi_{211}} \otimes \ket{\chi_+}+ b_{21-1\frac{1}{2}} (\mathcal{E}) \ket{\psi_{21-1}} \otimes \ket{\chi_+},  \\
 \ket{\psi_b} &=& \ket{\phi_{10\frac{1}{2}\frac{-1}{2}}}=\ket{\psi_{100}}\otimes\ket{\chi_-}, 
\end{eqnarray}
with
\begin{eqnarray}
         b_{200\frac{-1}{2}}(\mathcal{E}) &=& c_{20\frac{1}{2}\frac{-1}{2}}(\mathcal{E}) \\
        b_{210\frac{-1}{2}} (\mathcal{E}) &=& \sqrt\frac{1}{3}c_{21\frac{1}{2}\frac{-1}{2}}(\mathcal{E})+\sqrt\frac{2}{3}c_{21\frac{3}{2}\frac{-1}{2}}(\mathcal{E})  \\
       b_{211\frac{1}{2}} (\mathcal{E}) &=&  -\sqrt\frac{2}{3} c_{21\frac{1}{2}\frac{-1}{2}}(\mathcal{E}) \\
			 b_{21-1\frac{1}{2}} (\mathcal{E}) &=& \sqrt\frac{1}{3}c_{21\frac{3}{2}\frac{-1}{2}}(\mathcal{E})  
\end{eqnarray}
and ii) the transition between the same $\psi_a$ and $\psi^\prime_b=\phi_{10\frac{1}{2}\frac{1}{2}}=\psi_{100}\chi_+$.
Please note that, with the approximations described above, the ground state always has a fixed spin, while the excited state has contributions from both spin directions. In each case, the spin-changing transition elements vanish identically. 

As the host medium, we consider a glass with the real part of the permittivity $\epsilon_R=2.411$.
The imaginary part of glass permittivity $\epsilon_I$ is physically negligible. For demonstration purposes, we will consider the rather broad range $\epsilon_I\in (10^{-3},10^{-1})$. 

\begin{figure}\centering
\includegraphics[width=0.9\textwidth]{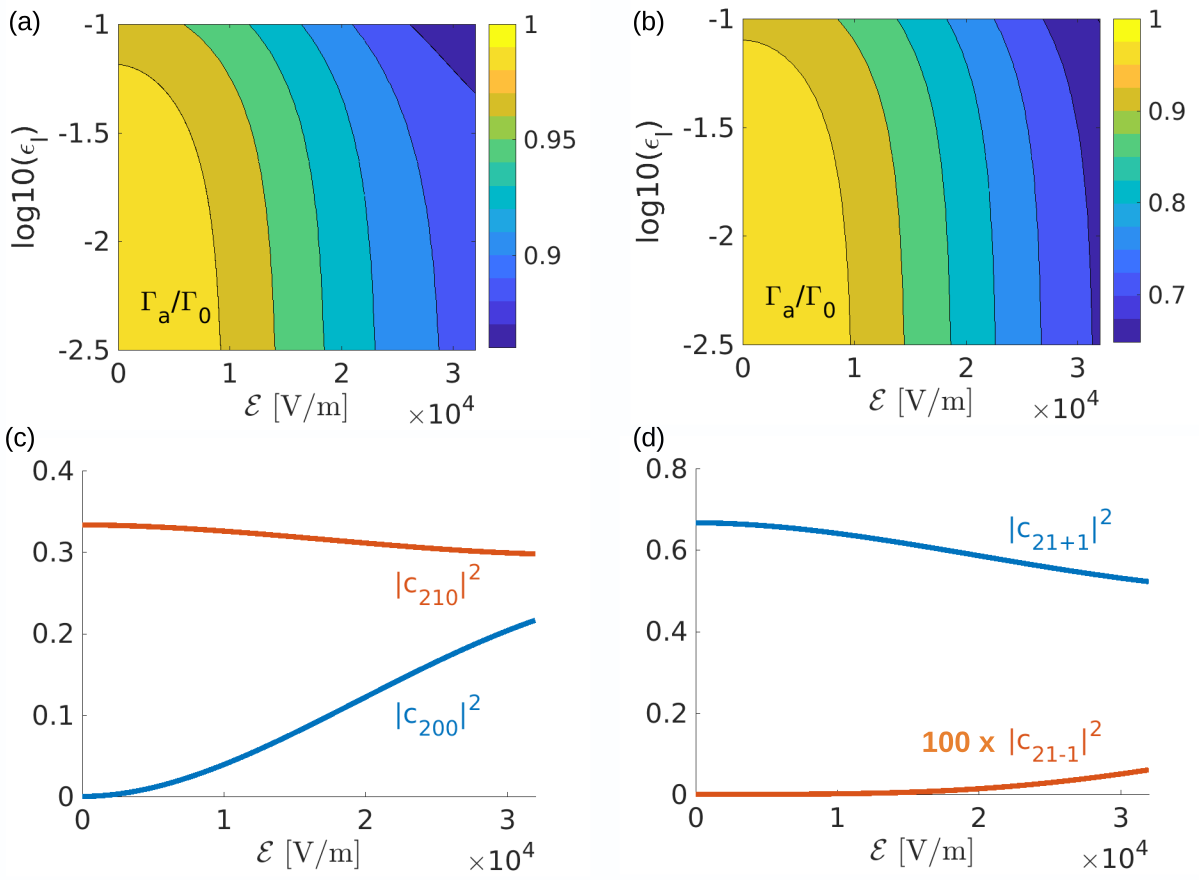}
\caption{Panels (a)-(b): Spontaneous emission rates from the lowest excited state $\ket{\psi_a}$ [Eq.~(\ref{eq:psi_hydrogen})] to the ground state corresponding to spin down (a) or up (b), of a hydrogen atom embedded in a medium with permittivity $\epsilon = 2.411+\ii\epsilon_I$ and subject to a static electric field $\mathcal{E}$ oriented along the quantization axis. The emission rate is normalized to the value at a vanishing electric field and $\epsilon_I=3.2\times 10^{-3}$.
Panels (c)-(d): Squared expansion coefficients of the state $\ket{\psi_a}$ in terms of bare hydrogen eigenstates. Please see the notation in Eq.~(\ref{eq:psi_hydrogen}).
}
\label{fig_example:hydrogen_emission1}
\end{figure}
    
We now evaluate $\Gamma_a$ applying the theory developed in Sec.~\ref{sec:emission} and leading to Eq.~(\ref{eq:Gamma}). 
The spontaneous emission rates for both transitions are displayed as functions of the external field $\mathcal{E}$ and the imaginary part of the permittivity $\epsilon_I$ in Fig.~\ref{fig_example:hydrogen_emission1}(a)-(b). 
As the asymmetry grows the transition rate is reduced in both cases, which is due to the increasing contribution of the ``dark'' component $\psi_{200}$ [Fig.~\ref{fig_example:hydrogen_emission1}(c,d)]: a transition between $\ket{\psi_{200}}$ and $\ket{\psi_{100}}$ is electric-dipole forbidden. 
We observe that the emission weakly depends on the absorption coefficient and slightly drops for larger values of the latter. 

We remark that, albeit these results have been obtained under the assumption of a homogeneous medium, which does not fully describe the physics of a system as small as a hydrogen atom, our analysis captures crucial information on the trends of the relevant physical quantities. 

	\subsection{Asymmetric quantum well}\label{asymmetry_large_coupled}
	
We evaluate here the decay rate for a semiconductor quantum well (QW). We consider the case of a symmetric and an asymmetric QW embedded in the same surrounding material. 
	
\begin{figure}\centering
		\includegraphics[width=0.6\textwidth]{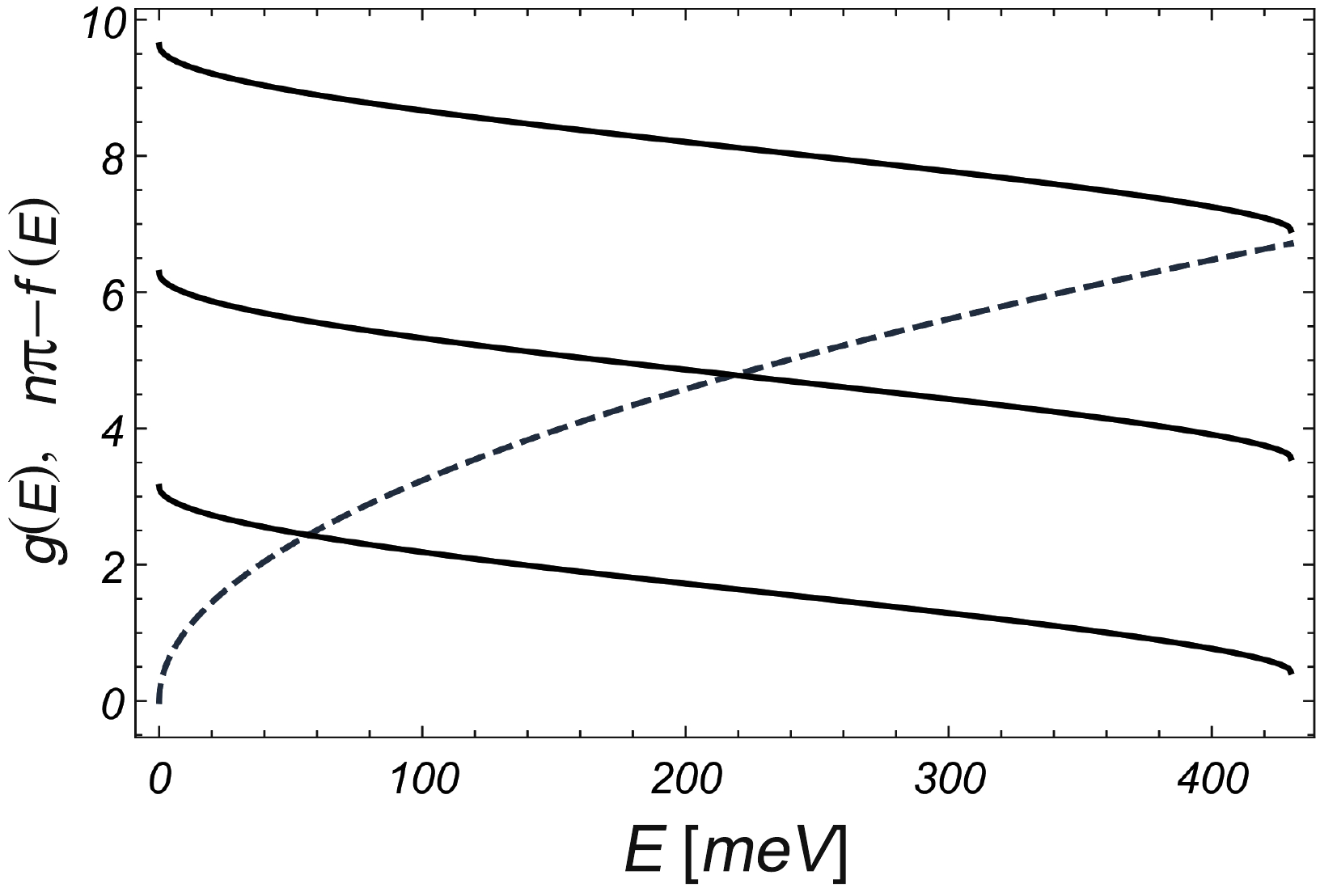}
		\caption{Graphical representation of Eq.~\ref{eq:intersection}. The dashed line represents the function $g(E)=a\sqrt{2mE/\hbar^2}$ [left-hand side of Eq.~(\ref{eq:intersection})], while the equally spaced solid lines represent $n\pi -f(E)$, where $f(E)=\arcsin{\sqrt{E/V_L}}+\arcsin{\sqrt{E/V_R}}$ [right-hand side of Eq.~(\ref{eq:intersection})], for $n=1,2,3$. The two intersections determine the eigenvalues $E_1$ and $E_2$. In this graph, the choice of the potentials $V_L$ and $V_R$ maximizes the width $a$ while keeping the number of bound energy levels equal to $2$.}\label{fig:twolevels}
	\end{figure}

	The considered  QW consists of Aluminium Indium Arsenide with different molar fractions ($\mathrm{Al}_x\mathrm{In}_{1-x}\mathrm{As}$ and $\mathrm{Al}_z\mathrm{In}_{1-z}\mathrm{As}$) and Gallium Indium Arsenide ($ \mathrm{Ga}_y\mathrm{In}_{1-y}\mathrm{As}$), with $x=0.46,y=0.48,z=0.47$, respectively. The well has a finite length $a$. The effective mass in the three regions is $m=0.043 m_e$, $ m=0.045 m_e$ and $m=0.078 m_e$, respectively, where $m_e$ is the electron mass. 
	By varying the molar fractions it is possible to modify the height of potential barriers $V_{L/R}$ on the left/right side of the QW, and consequently confine the electron along the $x$-direction with a potential~\cite{landau1981}
	\begin{equation}\label{eq:potentialQW}
	V\left(x\right)=
	\cases{V_{L} & for  $x< -a/2$\\
	0 & for  $-a/2\leq x\leq a/2$\\
	V_R & for  $x> a/2$.\\}
	\end{equation}
	Motion along the transverse $(y,z)$ directions is loosely bound; for simplicity, we will assume a weak harmonic confinement along those directions. The asymmetry of the system is related to the nonvanishing value of $V_R-V_L$ along the $x$-axis.
	In GaInAs, $V_{R}=520\,\mathrm{meV}$ and $V_{L}$ can be tuned with a
	sensitivity of $3\,\mathrm{meV}$~\cite{Faist1997,Vitiello2007}. The energy spectrum is determined by the following equation
	\begin{equation}
	a\sqrt{\frac{2mE}{\hbar^{2}}}=n\pi-\arcsin\sqrt{\frac{E}{V_{L}}}-\arcsin\sqrt{\frac{E}{V_{R}}},\label{eq:intersection}
	\end{equation}
	where $n=1,2$ correspond to the ground and the excited
	state, respectively. By tuning $V_{L}$, one can set the energy gap between the two lowest levels. We set the QW width $a$ to ensure the absence of a third bound level, as shown in Fig.~\ref{fig:twolevels}, and approximate our dynamics with the one of a two-level system.
	Typically, $a$ can be controlled with a precision of half a constant lattice $0.3\,\mathrm{nm}$~\cite{Faist1997}. The wavefunctions $\psi_n$, corresponding to the energy eigenvalues $E_n$, with $n\geq 1$, are given by
	\begin{equation}\label{eq:wavefunction}
	\Psi_n(x)=\begin{array}{c}
	c_n \cases{
	\sin(\delta_n)\ee^{\alpha_{nL}\left(x+\frac{a}{2}\right)} & for $ x < -\frac{a}{2}$\\
	\sin(\beta_n(x+\frac{a}{2})+\delta_n) & for $ -\frac{a}{2}\leq x\leq \frac{a}{2}$\\
	\sin(a\beta_n+\delta_n)\ee^{-\alpha_{nR}\left(x-\frac{a}{2}\right)} & for $ x>\frac{a}{2}$\\}\end{array}
	\end{equation}
	where $\alpha_{nR(nL)}=\sqrt{2m\left(V_{R(L)}-E_n\right)}/\hbar$, 
	 $\beta_n=\sqrt{2mE_n}/\hbar$,
	$\delta_n=\mathrm{arccot}\left(\alpha_{nL}/\beta_n\right)$
	and $c_n$ is a normalization constant. 
	
	The structure of the QW entails a trade-off
	between its width and the resonance wavelength $\lambda$ corresponding to the energy gap. The larger the width $a$, the larger the resonant wavelength, but the complex part of the permittivity becomes drastically smaller. This yields a medium that is practically transparent. For GaInAs,
	the relative permittivity is given by $\epsilon_R=11.638$ and $\epsilon_I=0.024082$ at the
	resonance energy gap of $161.917\, \mathrm{meV}$ determined by $V_{L}=430\,\mathrm{meV}$. 
	\begin{figure}\centering
		\includegraphics[width=0.6\textwidth]{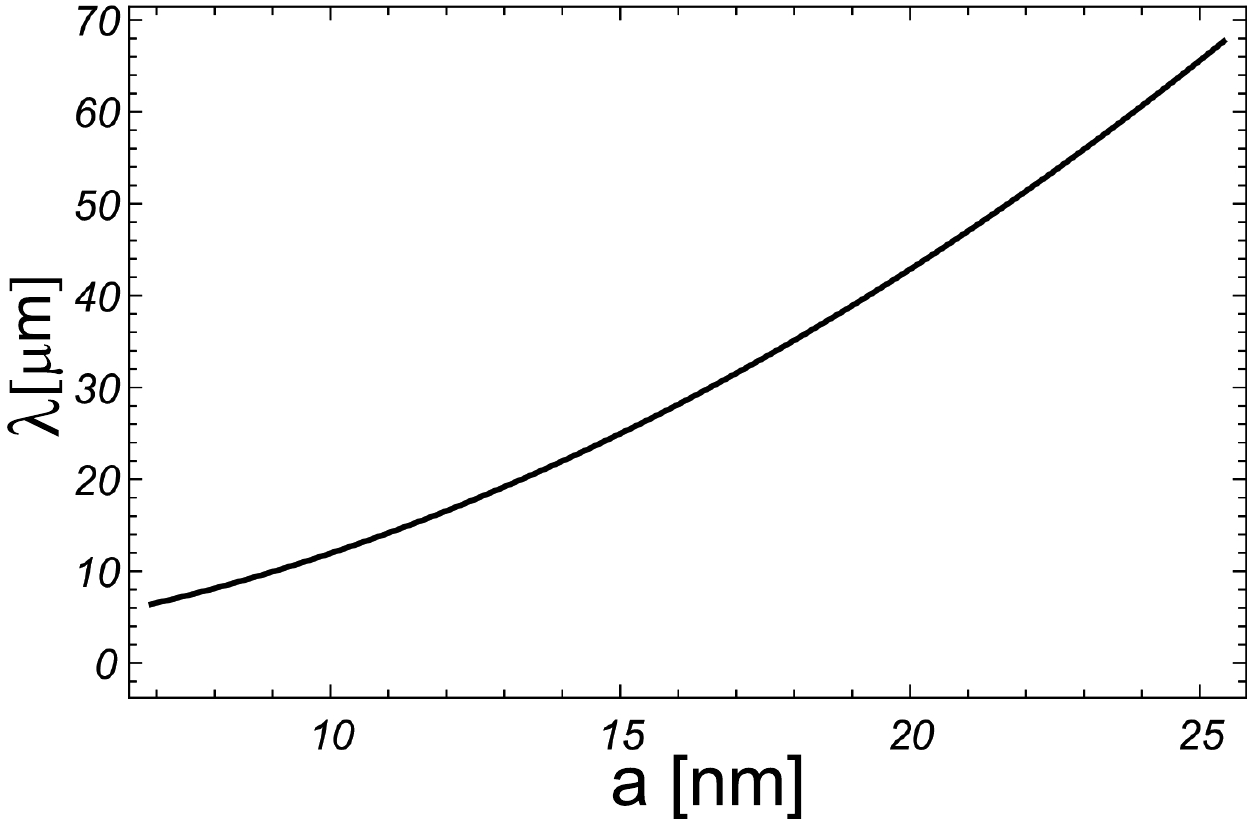}
		\caption{Resonant wavelength $\lambda= 2\pi\hbar c/(E_2-E_1)$ corresponding to the excitation energy from the ground state to the first excited state of the potential~\ref{eq:potentialQW}, as a function of the width of the quantum well for $V_{L}=V_R=430\,\mathrm{meV}$.}\label{fig:gap}
	\end{figure}

	 We consider the spontaneous transition between two states $a$ and $b$, characterized by the wavefunctions
	 \begin{equation}
	 \psi_{a}(\br)  = \Psi_2(-x) \frac{\ee^{-\frac{y^{2}+z^2}{4\sigma^{2}}}}{\sqrt{2\pi\sigma^{2}}}, \qquad \psi_{b}(\br)  = \Psi_1(-x) \frac{\ee^{-\frac{y^{2}+z^2}{4\sigma^{2}}}}{\sqrt{2\pi\sigma^{2}}},
	 \end{equation}

	 where the Gaussian part in the $(y,z)$ variables is related to the weak harmonic transverse confinement. Since the transverse wavefunction is the same for $a$ and $b$, the matrix elements of the dipole moment only have components along $x$:
	\begin{equation} 
		   e \left(\begin{array}{cc}
x_{aa} & x_{ab}\\
x_{ba} & x_{bb}
\end{array}\right)=
	    \left(\begin{array}{cc}
0.007 & 0.213\\
0.213 & 0.0153
\end{array}\right) e a_0,
	\end{equation}
	with $e=1.60\cdot 10^{-19}\,\mathrm{C}$ the electron charge and $a_0=5.29\cdot 10^{-11}\,\mathrm{m}$ the Bohr radius.

	\begin{figure}
	\centering
	\includegraphics[width=0.7\textwidth]{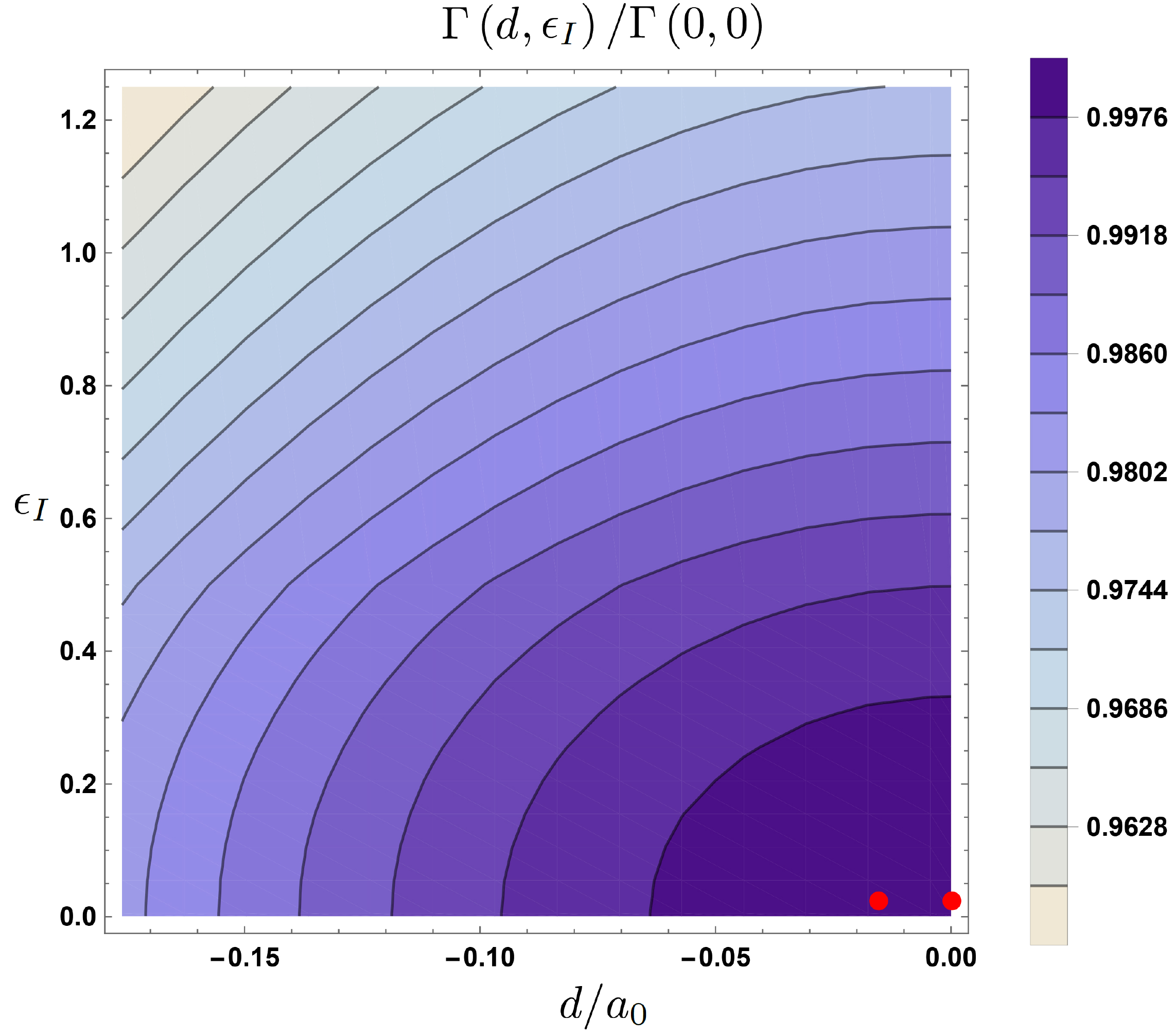}
	\caption{Ratio of the spontaneous emission rate $\Gamma$ over the symmetric and non dispersive one $\Gamma(0,0)$, as a function of $d/a_0$ and $\epsilon_I$ and at fixed $\epsilon_R=11.638$. The red dots represent the values of $\Gamma$ mentioned in the text for the approximated quantum well at $\epsilon_I=0.024$ with $d=0$ and $d=-0.0153\, e a_0$.}
	\label{fig:Gamma_eps_d}
	\end{figure}	
	
	\begin{figure}
	\centering
	\includegraphics[width=0.7\textwidth]{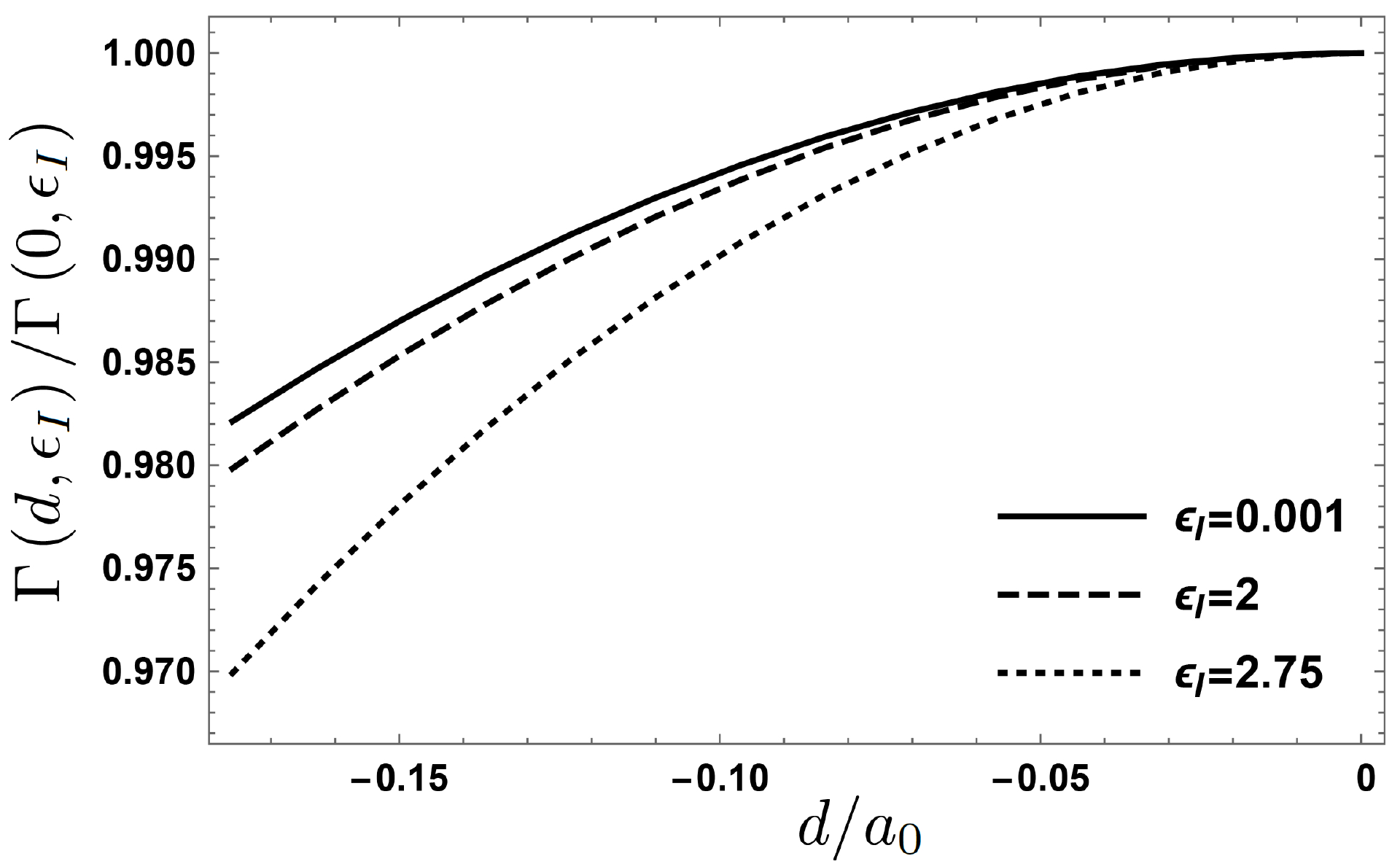}
	\caption{The ratio of the decay rate $\Gamma(d/a_0,\epsilon_I)/\Gamma(0,\epsilon_I)$ as a function of $d/a_0$ show the variation of the asymmetry for different dispersion media with $\epsilon_I=0.001$ (solid line), $\epsilon_I=2$ (dashed line) and $\epsilon_I=2.75$ (dotted line).}
	\label{fig:Gamma_eps_d2}
	\end{figure}

	Unlike  the case of the hydrogen atom, discussed in Sec.~\ref{sec:hydrogen_emission}, here both functions are characterized by a finite dipole moment. In order to highlight the specific effects of the average dipole moment of the two states [see Eq.~\ref{eq:rho}], we will consider an approximation in which the wavefunctions~(\ref{eq:wavefunction}) are replaced by harmonic oscillator eigenfunctions, both characterized by the same permanent dipole moment $\mu=e x_{aa}=e x_{bb}$. The frequency $\omega_{\mathrm{ho}}$ of the harmonic oscillator is fixed in such a way that $\hbar\omega_{\mathrm{ho}}$ matches the excitation energy from the ground state to the first excited state of the QW. The permanent dipole $\mu$ for the harmonic oscillator is obtained by shifting its wavefunctions along the $x$-axis such that $\mu=0.0153\,\, e  a_0$. Hence, 
	\begin{equation}
	\Psi_{1}\left(-x\right)\simeq \frac{\ee^{-\frac{\left(x+x_{aa}\right)^{2}}{4\sigma_{x}^{2}}}}{\sqrt[4]{2\pi\sigma_{x}^{2}}},\quad
	\Psi_{2}\left(-x\right)\simeq \frac{x+x_{aa}}{\sigma_x}\frac{\ee^{-\frac{\left(x+d\right)^{2}}{4\sigma_{x}^{2}}}}{\sqrt[4]{2\pi\sigma_{x}^{2}}}\label{eq:psib-1}
	\end{equation}
	with $\sigma_{x}^{2}=\hbar/\left(2m\omega_{\mathrm{ho}}\right)$.
    We obtain the decay rates $\Gamma(d=-0.0153\, e a_0)=3.29708\cdot 10^{15}\, \mathrm{s^{-1}}$ and $\Gamma(d=0)=3.29773\cdot 10^{15}\, \mathrm{s^{-1}}$, yielding a $0.02\%$ increase of the asymmetric case compared to the symmetric one, of the same order of the ratio $d/(ea_0)$ [see Eqs.~(\ref{eq:Fabappr}-\ref{eq:Gabappr}) and comments thereto]. 
	In Fig.~\ref{fig:Gamma_eps_d}, we show the results for the spontaneous emission rate with varying $d$ and $\epsilon_I$, at fixed $\epsilon_R=11.638$. In vacuum ($\epsilon\to1$), the relative contribution of the asymmetry to the total decay rate becomes less relevant. Furthermore, to highlight the specific effect of a finite dipole moment, we show in Fig.~\ref{fig:Gamma_eps_d2} the ratio between the value of the decay rate as a function of $d$ and its value for $d=0$, corresponding to a fixed $\epsilon_I$.

	\section{Conclusions}\label{sec:conclusions}
	
	We studied the dynamics of a charged system coupled to a medium-assisted electric field, beyond the point-dipole approximation, highlighting the role played by the finite size of the system, the dispersion and absorption by the medium and the spatial asymmetries. The analysis  focused on the determination of the decay rates and energy shifts of the bound states of an ``atomic'' system, which have been obtained under general assumptions. The most important among these assumptions is the hypothesis of homogeneous and isotropic media. We also discussed how to extend the theory to more general situations. 
	
	The   obtained results were applied to two test-beds: a microscopic one, represented by a hydrogen atom subject to a uniform electric field, and a mesoscopic one, consisting of a quasi-electron in a semiconductor quantum well. In both cases, we have obtained  the decay rates as functions of the asymmetry of the system and the absorption of the medium, showing that asymmetry can yield small but detectable deviations with respect to the symmetric case.
	
	Future research will be devoted to a thorough treatment of medium inhomogeneity and anisotropy and, in particular, to the inclusion of effects due to the medium granularity, which implies a further lenght scale and momentum cutoff, competing with those related to  the atomic system size.

	\ack
		PF, FVP, SP and GS were partially supported by Istituto Nazionale di Fisica Nucleare (INFN) through the project ``QUANTUM''. FVP acknowledges support by MIUR via PRIN 2017 (Progetto di Ricerca di Interesse Nazionale), project QUSHIP (2017SRNBRK) and via PON ARS $01\_00141$ CLOSE.  PF is partially supported by the Italian National Group of Mathematical Physics (GNFM-INdAM).
		KS acknowledges the National Science Centre, Poland grant number 2018/31/D/ST3/01487. GS was supported by the PROM  Project, funded by Polish National Agency for Academic Exchange ``NAWA'',   agreement no.\ PPI/PRO/2018/1/00016/U/001, and the Torun Astrophysics/Physics Summer Program TAPS 2018.
PF, FVP and SP are partially supported by Regione Puglia and by QuantERA ERA-NET Cofund in Quantum Technologies (GA No.\ 731473), project PACE-IN. 

\section*{References}

%\bibliographystyle{ieeetr}
%\bibliography{asymmetricBib}

\end{document}